# Modularized Neural Network Incorporating Physical Priors for Smart Building Control

Accuracy or Consistency?

Zixin Jiang
Dept. of Mech. & Aero. Eng.
Syracuse University
Syracuse NY USA
zjiang19@syr.edu

Bing Dong
Dept. of Mech. & Aero. Eng.
Syracuse University
Syracuse NY USA
bidong@syr.edu

## ABSTRACT

Model predictive control can achieve significant energy savings, offer grid flexibility, and mitigate carbon emissions. However, the challenge of identifying individual control-oriented building dynamic models limits large-scale real-world applications. To address this issue, this study proposed a Modularized Neural Network Incorporating Physical Priors (ModNN), capable of establishing a control-oriented and physical-consistent building dynamic model within minutes without substantial modeling effort. This is also the first study to evaluate the physical consistency of a given data-driven model both qualitatively and quantitively. We compared the physical consistency of a classical Long Short-Term Memory (LSTM) model and our ModNN. The ModNN strictly satisfies physical constraints, whereas the LSTM model learned contradictory system dynamics. Additionally, we compared their control performance on an EnergyPlus virtual testbed. While the LSTM model demonstrated slightly better prediction accuracy in dynamic modeling, it failed in control optimization, resulting in an 89°C-h temperature violation, whereas the ModNN showed only a 0.57°C-h violation and achieved up to a 78% peak load reduction. Our findings highlight the importance of incorporating physics priors into data-driven models and provide a promising solution for future smart building control optimization. Furthermore, the proposed evaluation framework defines two physical consistency indicators, providing guidelines for selecting and testing control-oriented, data-driven building dynamic models.

## CCS CONCEPTS

**ACM Reference format:**

## 1 Introduction

Buildings account for 30% of global final energy consumption and 27% of global energy-related emissions [1]. Among various building energy consumers, Heating, Ventilation, and Air-Conditioning (HVAC) systems account for more than half of the used energy [2]. However, 40% of this energy is wasted due to inappropriate HVAC control, mismatched operation schedules, and other inefficiencies [3]. Therefore, developing advanced HVAC control strategies is crucial for reducing building energy consumption, mitigating global warming, and promoting carbon neutrality.

One effective measure to achieve this is model predictive control (MPC). By minimizing a cost function based on predicted system dynamics and disturbances such as weather and occupancy, MPC can achieve significant energy savings and enhance grid flexibility in buildings [4,5,6,7]. Despite the attractive energy-saving benefits of MPC, its widespread adoption faces significant challenges. The major barrier is the difficulty of developing control-oriented models. Since MPC solves optimization problems based on system dynamics, the model must accurately capture the system's response to control inputs and disturbances. However, developing and maintaining such a model requires detailed building information, expert knowledge, substantial modeling efforts, and customized case-by-case calibrations [8,9]. This challenge potentially hinders its large-scale commercial applications.

Nowadays, the rapid development of data-driven techniques and the increasing accessibility of data make data-driven predictive control one of the most promising approaches to address this scalability challenge due to its powerful approximation ability of nonlinear processes [10]. A data-driven model can listen to real data from existing systems and interfaces, capturing the mathematical relationship between building operation data and energy consumption without substantial modeling efforts. Compared with traditional white-box or gray-box models, data-driven models demonstrate better accuracy and higher efficiency [11,13], providing a feasible solution for large-scale MPC implementation. For example, Lee et al. [17] developed a simplified autoregressive with exogenous inputs (ARX) model for predicting the indoor temperature. The proposed MPC framework achieved up to approximately 12% heating energy savings. Cotrufo et al. [18] adopted MPC based on five data-driven models: 1) artificial neural networks (ANN), 2) Gauss process regressions (GPR), 3) support vector machines (SVM), 4) decision trees (DT), and 5) random forests (RF). The results shown that the GPR models was the most accurate and it can reduce 22% of natural gas consumption and greenhouse gas emissions. Stoffel et al. [20] implemented MPC on two case buildings using three data-driven models: ANN, GPR, and linear regression (LR). The authors recommend the LR because it requires less effort to set up, despite ANN showing the higher prediction accuracy. Li et al. [19] implemented MPC on an encoder-decoder based recurrent neural network and achieved 4% to 7% energy savings. Among these models, deep learning-based technology demonstrates better prediction accuracy. For instance, Mtibaa et al. [21] demonstrate that LSTM outperforms ANN in HVAC systems by as much as 50% in accuracy. Wang et al. [22] compared 12 data-driven models and the result shown that LSTM performs best in short-term prediction.

While data-driven models achieve a reasonable estimation of indoor temperature dynamics within 5-7% mean absolute percentage error (MAPE) based on literature review [15], they are subject to limited interoperability and generalization ability [14], and they are highly sensitive to data quality and quantity [13]. Especially in the building control field, most buildings operate according to limited control modes, weather conditions, and setpoints. The space air temperature is typically maintained within a comfort zone constrained by one or two fixed setpoints [13]. The limited training dataset could cause overfitting, meaning that a data-driven model has poor ability to forecast out-of-sample data (unseen situations). Another significant barrier to real-world data-driven MPC implementation is the lack of physical consistency guarantees for data-driven models. For example, as shown in the work by Di Natale et al. [15], the authors compared a physically consistent neural network with a conventional Long Short-Term Memory network (LSTM). Despite the LSTM demonstrating superior prediction accuracy, it failed to capture the underlying physics, such as the impact of heating and cooling on temperature.

Consequently, to leverage the powerful modeling accuracy of data-driven method without losing physical consistency guarantee, the state-of-the-art method is to incorporate physical priors into deep neural networks, and there are three typical methods to achieve this [16]:

1) Adjusting model structures: Researchers [15, 16, 24, 26, 27] adjust neural network structures based on prior knowledge to ground them in underlying physical laws, thereby improving the model's generalization ability.
2) Adding physical constraints to model parameters: Researchers [15, 16, 26, 27, 28, 29] constrain model parameters to ensure correct model responses to inputs.
3) Customizing loss functions: Researchers [23, 25] introduce a physical loss term calculated by a physical model to make the network adhere to the underlying physics.

Although incorporating physical priors into neural networks demonstrates higher accuracy, enhanced data efficiency, and better generalization ability compared to traditional purely data-driven models, there is a lack of research evaluating the control performance difference between classical data-driven methods, such as LSTM, with physics-inspired neural networks. Another research gap is the absence of performance indicators for selecting the appropriate control-oriented model. Most data-driven methods primarily consider accuracy metrics such as mean squared error (MSE) or mean absolute error (MAE), which are insufficient for control purposes.

To address the aforementioned challenges, we summarize our contributions as follows:
- We proposed a modularized neural network incorporating physical priors, which is physically consistent and control oriented.
- We compared the control performance difference of an LSTM-based building dynamic model with our ModNN on an EnergyPlus virtual testbed.
- We developed an evaluation framework to evaluate the physical consistency of data-driven models based on maximum mean discrepancy (MMD) and temperature response violation (TRV). This is also the first quantification study of physical consistency in building energy modeling filed which can benefit future researcher to select the appropriate data-driven building energy models.
- We developed a control law neural network which can solve highly nonlinear optimization problems caused by complex dynamic models in real time.

The paper is organized as follows: Section 2 presents the overall methodology, including the co-simulation setup, model structure, evaluation matrix, and optimization formulation. Section 3 compares the accuracy, consistency, and control performance of the proposed ModNN and LSTM models. Section 4 discusses the trade-off between consistency and accuracy. Section 5 pictures the limitation and future studies. Section 6 provides the conclusions, and Section 7 lists the references.

## 2  Methodology

This section introduces the overall methodology as shown in Figure 1. We begin with a detailed introduction of the EnergyPlus virtual testbed in Section 2.1. Subsequently, we describe the proposed ModNN structure in Section 2.2. Next, we display the physical consistency evaluation framework in Section 2.3 and finally, Section 2.4 depicts the energy optimization formulation by a control law neural network.

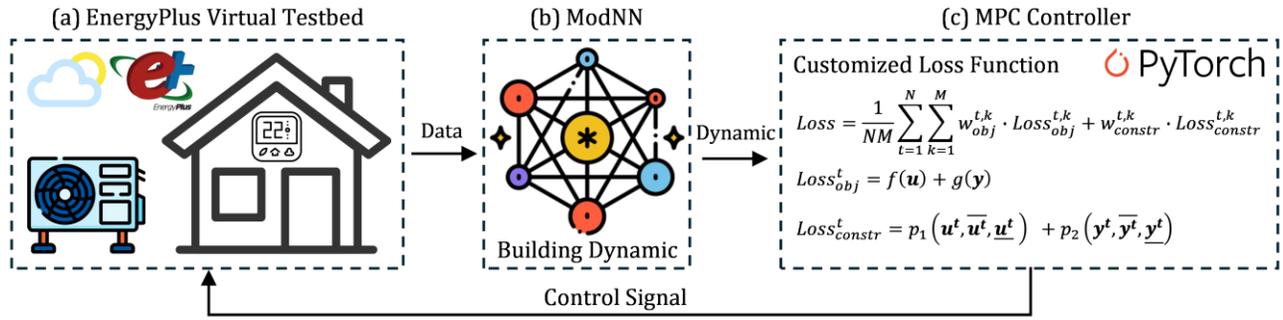

**Figure 1** Overall diagram of proposed EnegyPlus-ModNN-MPC co-simulation framework

## 2.1 EnergyPlus Virtual Testbed

We conduct our simulation on EnergyPlus based on a single-family prototype building [30] developed by Pacific Northwest National Laboratory. We use EnergyPlus Runtime API to mimic the data collection and HVAC system control process. This API class allows a client to interface with EnergyPlus at runtime, enabling data sensing and actuation during a running simulation.

**Table 1 Simulation settings for EnergyPlus virtual testbed**

| Parameters | Values |
| --- | --- |
| Weather Condition | Denver, Climate 5, Cool Dry |
| Timestep | 15 Minutes |
| Setpoint (occupied) | 24 °C |
| Setpoint (unoccupied) | 32 °C |
| Baseline | On-Off control with deadband |
| Deadband | 0.5 °C |
| Depart Time | $\mathcal{U}(7:00, 10:00)$ |
| Arrive Time | $\mathcal{U}(16:00, 20:00)$ |
| Supply Air Temperature | 13 °C |
| Supply Air Flow Rate | 0 to 0.16 m³/s |
| Simulation Period | Summer (June 1st to Aug 31st) |
| Peak Hour | 15:00 to 18:00 |

At each time step (15 minutes), we collect data on outdoor air temperature, solar radiation, occupancy level, HVAC power, and space air temperature via EnergyPlus variable handle. And we send control signal via actuator handle to adjust the supply air flow rate according to the control policy. Detailed simulation settings can be found in Table 1.

## 2.2 Modularized Neural Network Incorporating Physical Priors

In this subsection, we introduce how we incorporate physical priors into our ModNN from the following aspects:

### 2.2.1 Physics-inspired modularization

Inspired by the heat balance equation as shown in Eq. 1, where $m_{air}$ is the mass of space air (kg), $c_{air}$ is the specific heat of space air (J/(kg·°C)), $T_{air}$ is the space air temperature (°C), $\dot{q}_{conv,sur}$ is the convective heat transfer from interior surfaces (W), $\dot{q}_{conv,int}$ is the convective heat transfer from light, equipment, and occupants (W), $\dot{q}_{infiltration}$ is the heat transfer through infiltration (W), and $\dot{q}_{HVAC}$ is the heat transfer from HVAC system (W). We aggregate these terms from a data-driven perspective, using three functions $f_{ext}(\circ)$, $f_{int}(\circ)$, and $f_{HVAC}(\circ)$ to represent the heat transfer from external, internal, and HVAC systems, respectively. We developed separate neural networks to estimate each of these heat transfer terms.

$$m_{air} \cdot c_{air} \cdot \frac{dT_{air}}{dt} = \dot{q}_{conv,sur} + \dot{q}_{conv,int} + \dot{q}_{infiltration} + \dot{q}_{HVAC} = f_{ext}(\circ) + f_{int}(\circ) + f_{HVAC}(\circ) \quad 1$$

Unlike classical neural network structures that process all inputs collectively—feeding all related features into one model directly without any guidance, thereby losing physical interpretability—our approach refines the model based on heat balance principles. This allows each module within the model to be physically meaningful. For example, the internal heat gain in buildings typically originates from three sources: (1) metabolic heat generated by occupants, (2) heat from electrical equipment and appliances, and (3) heat from lighting. Thus, the internal heat gain module uses 'occupancy', 'time', and 'space air temperature' as inputs to predict the internal heat gain. A more detailed description of the model inputs can be found in Table 2. And a general introduction of the model structure can be found in Figure 2.

### 2.2.2 Physics-inspired model structure

Inspired by the state-space formulation, we developed a sequence-to-sequence encoder–decoder model structure to represent building thermos dynamics. First, an encoder is designed to extract historical information. For example, it is used to transform the heat stored in building envelopes and furniture owing to thermal inertia into a high-dimensional latent vector. Then, a current cell is designed to take a measurement of the current time step, and the measurement is combined with the latent hidden vector form the encoder. The purpose is to eliminate the accumulated modeling errors and provide a more accurate initial points for decoder. Finally, a decoder is designed to predict the system responses based on future inputs, using the initial state from the encoder and current cell as an initial point, the decoder can generate a sequential system response according to future disturbances and system inputs.

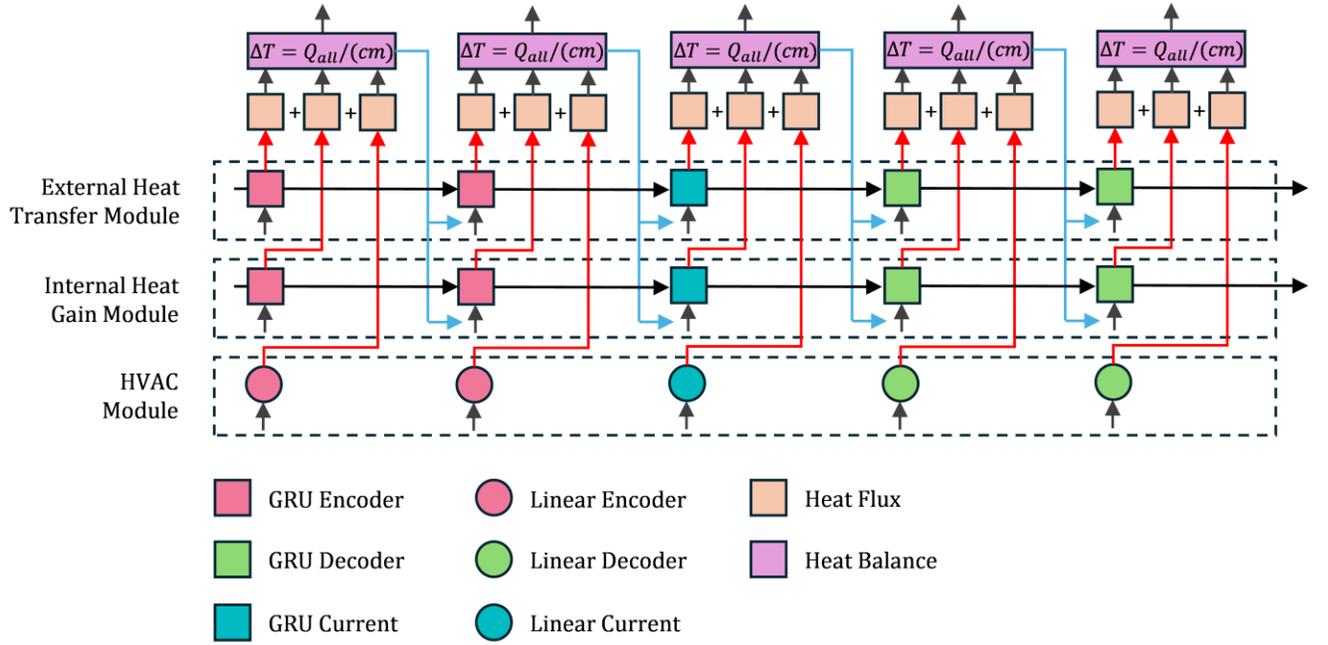

**Figure 2 Overall structure of proposed ModNN**

**Table 2 Model inputs for ModNN**

| Inputs | Module | Outputs |
|---|---|---|
| Outdoor Air Temperature | External Heat Transfer Module (Seq2Seq Gru) | Latent Heat Flux from External |
| Solar Radiation | | |
| Time of A Day | | |
| Occupancy | Internal Heat Transfer Module (Seq2Seq Gru) | Latent Heat Flux from Internal |
| Time of A Day | | |
| HVAC Power | HVAC Module (Seq2Seq Linear) | Latent Heat Flux from HVAC |
| Latent Heat Flux from External, Internal and HVAC | Heat Balance Module (Seq2Seq Linear) | Temperature Change |

*2.2.3 Physics-inspired model constraints*

To ensure proposed ModNN responds appropriately to a given model input, we add physical consistency constraints to model parameters to ground it with physical principles. For example, the indoor air temperature should decrease with an increasing HVAC cooling load, and vice versa for heating. This consistency can be ensured by forcing the partial derivative of the model output to be positive with respect to its input as shown in Eq. 2:

$$\frac{\partial y_t}{\partial u_t^{HVAC}} > 0 \qquad 2$$

Where $y_t$ represents the indoor temperature at timestep t, it can be calculated by Eq. 3, and $u_t^{HVAC}$ represents the HVAC power at timestep t.

$$\begin{aligned} y_t &= y_{t-1} + FC(Q_t^{all}) \\ &= y_{t-1} + FC(Q_t^{ext} + Q_t^{int} + Q_t^{HVAC}) \\ &= y_{t-1} + FC(Q_t^{ext} + Q_t^{int} + FC(u_t^{HVAC})) \end{aligned} \qquad 3$$

Where $FC$ represents the fully connected neural network, $Q_t^*$ represents different heat flux term as shown in Figure 2 and Eq. 1. To satisfy the constraints from Eq. 2, the weight of each fully connect neural network layer has to be strictly positive. Another very import trick is that these two fully connected neural networks cannot use the rectified linear unit (ReLU) activation function; otherwise, the outputs for cooling and heating will both be clipped to non-negative values.

## 2.3 Physical Consistency Evaluation

In this section, we introduced two indicators to evaluate the physical consistency of proposed ModNN.

*2.3.1 Temperature response violation*

According to the underlying heat balance laws, space air temperature should increase with increased heating or decreased cooling, and vice versa. At each time step, we inject maximum cooling (blue line) or turn off the cooling (red line) instead of using the original cooling input (green line) to test if the model responds correctly as shown in Figure 3. If we introduce more cooling to the space, but the space air temperature increases compared to the predicted temperature based on the original cooling, we define the summation of this incorrect temperature violation as TRV⁻. Similarly, if the temperature decreases with less cooling, we define the summation of this incorrect temperature violation as TRV⁺. Calculation details can be found in Eq.4:

$$TRV^+ = sum(\min(T_{up} - T_{pred}), 0)$$
$$TRV^- = sum(\min(T_{pred} - T_{down}), 0) \qquad 4$$

Where $T_{up}$ is the temperature with less cooling or more heating and $T_{down}$ is the temperature with more cooling or less heating.

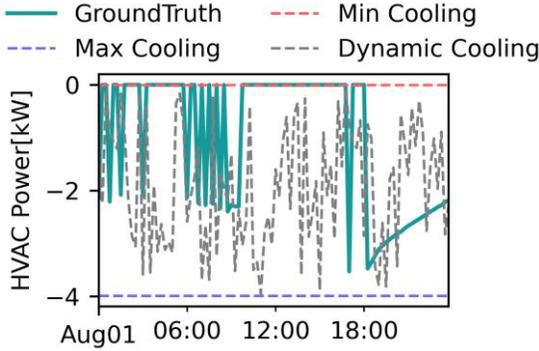

**Figure 3 Cooling load for sanity check**

*2.3.2 Maximum mean discrepancy*

The aforementioned TRV is used to check if the model responds correctly, but it cannot indicate the accuracy of the model's response. For example, if model 1 and 2 both predict an increased space air temperature with less cooling, TRV cannot determine which model is closer to the real system. Therefore, we use another metric to quantify how consistent the data-driven model is compared to the real dynamic system.

Maximum mean discrepancy (MMD) is a distance on the space of probability measures which has found numerous applications in machine learning and nonpara- metric testing [32]. It measures the distance between two sets of samples by taking the maximum difference in sample averages over a kernel function $\varphi$ [31]. Where $\varphi$ is a feature map $\varphi: \mathcal{X} \rightarrow \mathcal{H}$ and $\mathcal{H}$ is Hilbert space. $\mathcal{K}(x,y) = <\varphi(x), \varphi(y)>_\mathcal{H}$. And in general, MMD can be calculated by Eq. 5:

$$MMD^2(P, Q) = \left\Vert \mathbb{E}_{X \sim P}[\varphi(X)] - \mathbb{E}_{Y \sim Q}[\varphi(Y)] \right\Vert_\mathcal{H}^2$$
$$= \mathbb{E}_{X,X' \sim P}\mathcal{K}(X, X')$$
$$+ \mathbb{E}_{Y,Y' \sim Q}\mathcal{K}(Y, Y') \qquad 5$$
$$- 2\mathbb{E}_{X \sim P, Y \sim Q}\mathcal{K}(X, Y)$$

In this study, we select the widely used Gaussian kernel which can be calculated by Eq. 6:

$$\mathcal{K}(X, Y) = \exp\left(-\frac{1}{2\sigma^2}\Vert X - Y \Vert^2\right) \qquad 6$$

Where $X, Y$ is the sample from dataset P and Q. To obtain P and Q, we first randomly generated a testing HVAC load sequence as shown in Figure 3 (gray line), passed it to the model, calculated the one-step HVAC power change and the corresponding temperature change, this two-column dataset is denoted as dataset P. Similarly, we calculated the one-step HVAC power change and its corresponding temperature change from the raw data, resulting in dataset Q. Finally, we compared the similarity of datasets P and Q using MMD as introduced above.

## 2.4 Energy Optimization

Data-driven models are often complex, highly nonlinear, and hard to express explicitly. These challenges make traditional optimization solvers difficult to find the optimal solution due to local minima issues or unaffordable computation time. To overcome this challenge, we use the stochastic gradient-based solver from PyTorch, which is well-designed for neural network optimization, to solve our energy optimization problem.

First, we train our ModNN and freeze its parameters. We then integrate this model with a control law neural network. At each time step, ModNN predicts the system dynamics (future temperature sequence in this case) based on the control input and disturbances (such as weather and occupancy) under prediction horizons. The control signals from the control law neural network, where we replaced the mean square error loss to an MPC-like loss function as shown in Eq. 7:

$$Loss = \frac{1}{NM} \sum_{t=1}^{N} \sum_{k=1}^{M} w_{obj}^{t,k} \cdot Loss_{obj}^{t,k} + w_{constr}^{t,k} \cdot Loss_{constr}^{t,k} \qquad 7$$

Where t is the time index and k are the index of different loss terms. $Loss_{obj}^{t,k}$ refers to objective function such as load or energy cost and $Loss_{constr}^{t,k}$ refers to the constraint violation penalty such as temperature violation. For example, for a classical MPC problem that aims to maintain temperate while minimizing energy consumption, we can formulate the problem as shown in Eq. 8:

$$Loss = f(\mathbf{u}) + p_1\left(\mathbf{u}^t, \overline{\mathbf{u}^t}, \underline{\mathbf{u}^t}\right) + p_2\left(\mathbf{y}^t, \overline{\mathbf{y}^t}, \underline{\mathbf{y}^t}\right)$$
$$f(\mathbf{u}) = \sum_{t=1}^{M} \left(\frac{price_t \cdot u_t^{hvac}}{COP_t}\right)^2 \qquad 8$$

$$p_1\left(\boldsymbol{u^t},\overline{\boldsymbol{u^t}},\underline{\boldsymbol{u^t}}\right) = \sum_{t=1}^{M}\left[\left(\min\left\{\boldsymbol{0},\left(\boldsymbol{u}_t^{hvac}-\underline{\boldsymbol{u}_t^{low}}\right)\right\}\right)^2\right.$$
$$\left.+\left(\max\left\{\boldsymbol{0},\left(\boldsymbol{u}_t^{hvac}-\overline{\boldsymbol{u}_t^{high}}\right)\right\}\right)^2\right]$$

$$p_2\left(\boldsymbol{y^t},\overline{\boldsymbol{y^t}},\underline{\boldsymbol{y^t}}\right) = \sum_{t=1}^{M}\left[\left(\min\left\{\boldsymbol{0},\left(\boldsymbol{y}_t^{zone}-\underline{\boldsymbol{y}_t^{spt_{low}}}\right)\right\}\right)^2\right.$$
$$\left.+\left(\max\left\{\boldsymbol{0},\left(\boldsymbol{y}_t^{zone}-\overline{\boldsymbol{y}_t^{spt_{high}}}\right)\right\}\right)^2\right]$$

## 3 Results

### 3.1 Model Prediction Performance and Temperature Response Violation Evaluation

We compared the prediction accuracy and RVE of the LSTM and ModNN models as shown in Figure 4 and Figure 5. The gray lines represent the space air temperature predictions from the LSTM and ModNN models, respectively. Each line describes the temperature trajectory one day ahead (96-time steps), updated every 15 minutes. The green line indicates the ground truth results from EnergyPlus. The red and blue lines show the predicted space air temperature under minimal and maximum HVAC cooling loads (as shown before in Figure 3).

It is evident that both LSTM and ModNN align closely with the ground truth. We use mean absolute error (MAE) and MAPE as key indicators to evaluate model prediction accuracy, where LSTM outperforming ModNN. The MAE and MAPE of the LSTM model are 0.36 °C and 1.32%, respectively, compared to the ModNN model, which has an MAE and MAPE of 0.75 °C and 2.87%.

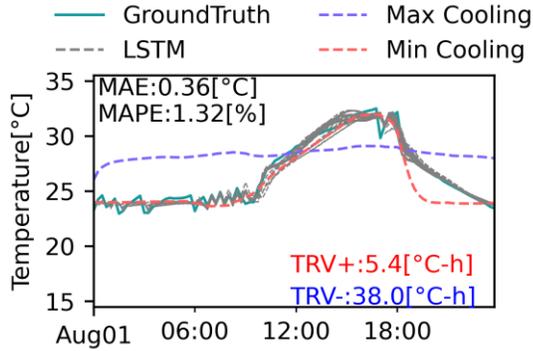

Figure 4 LSTM model performance

However, from a consistency perspective, the TRV+ and TRV- of the LSTM model are 5.4 °C-h and 38 °C-h, respectively. This indicates that despite the LSTM model's high alignment with the ground truth, it does not learn the correct system response. As shown by the blue line in Figure 4, a higher cooling load even increases space air temperature, and the temperature barely changes after the HVAC is turned off. In contrast, the ModNN model is always physically consistent with 0 °C-h TRVs, resulting in a more accurate system response. As illustrated in Figure 5, the space air temperature decreases after increasing cooling and rises in the afternoon due to hot outdoor air temperature and solar radiation.

The space air temperature increases after turning off the HVAC but follows a similar trend to the baseline during the afternoon since the original HVAC load is also 0 during that period.

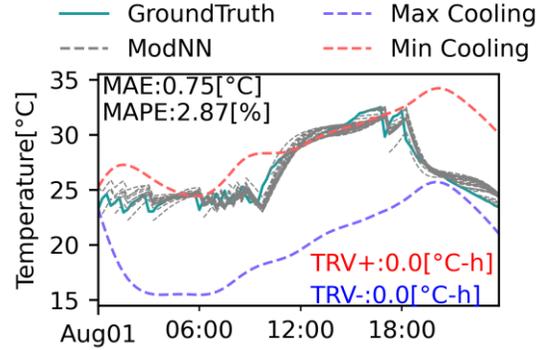

Figure 5 ModNN model performance

### 3.2 Physical Consistency Evaluation Based on Jacobian Matrix

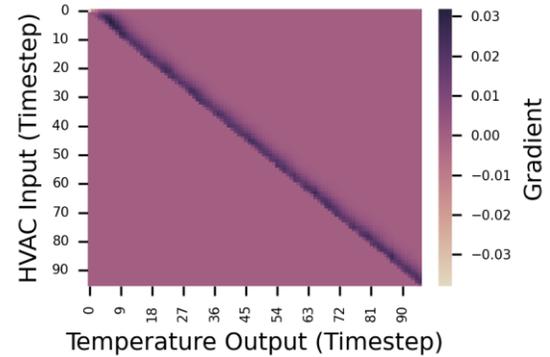

Figure 6 Jacobian Matrix of LSTM

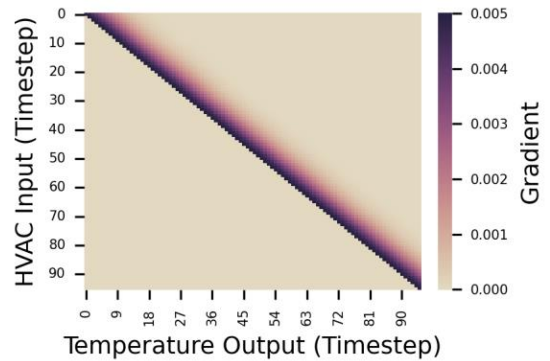

Figure 7 Jacobian Matrix of ModNN

We computed the Jacobian matrix of space air temperature given the HVAC inputs. As shown in Figure 6, we present the Jacobian matrix of the LSTM model, where the Y-axis represents the 96-timestep prediction inputs (HVAC power) and the X-axis represents the corresponding outputs (temperature). The color map indicates the gradient of each output with respect to the inputs. The Jacobian matrix of the LSTM model is not always positive, indicating that it cannot ensure physical consistency (violates Eq. 2). The Jacobian matrix of ModNN is shown in Figure 7, since we incorporated physical consistency constraints, the gradient of proposed ModNN is strictly positive.

## 3.3 Loss Evaluation

Figure 8 depicts the loss curve of the LSTM model. The mean square error (MSE) loss decreases rapidly on both the training dataset and validation dataset, as shown by the red and green lines. However, the TRV loss is unstable and consistently higher than 0, indicating that physical consistency cannot be guaranteed, and the response is incorrect. Figure 9 displays the loss curve of the ModNN model. Here, the initial MSE loss is higher than LSTM model, but it decreases with more training epochs. And the TRV loss remains at 0, demonstrating that physical consistency is consistently maintained.

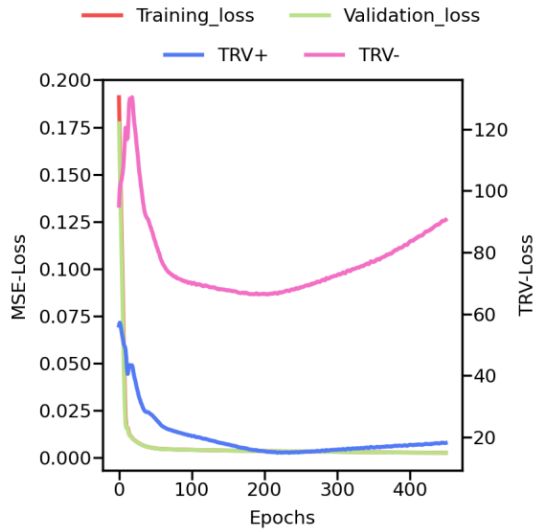

Figure 8 Loss of LSTM

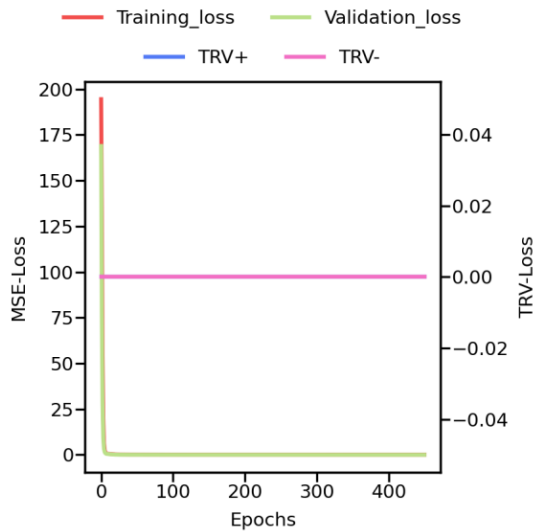

Figure 9 Loss of ModNN

## 3.4 Physical Consistency Evaluation Based on Evaluation Maximum Mean Discrepancy

We compared the Maximum Mean Discrepancy (MMD) of the LSTM and ModNN models, as shown in Figure 10. The X-axis represents one-step HVAC load change (negative values indicate more cooling, while positive values indicate less cooling), and the Y-axis represents the corresponding space air temperature change. Each scatter point is collected based on different weather and occupancy conditions.

From the black points (ground truth), we observe that space air temperature shows a clear decreasing trend with increased cooling load intake. The ModNN model demonstrates a similar trend, as shown by the blue points. However, for the LSTM model, as depicted by the red points, the space air temperature increases with more cooling. There is a clear difference between the distribution of the two datasets (black contour plot and red contour plot). The MMD of the LSTM model to the ground truth is 0.14, which is much higher than the MMD of the ModNN model to the ground truth (0.05).

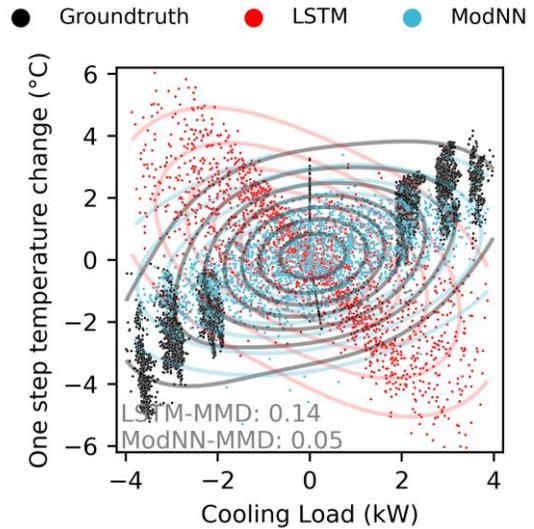

Figure 10 Maximum mean discrepancy of LSTM and ModNN

## 3.5 Control Performance Compare between LSTM and ModNN

We compared the control performance of ModNN, LSTM, and the baseline on an example day, as shown in Figure 11. For the on-off baseline control, represented by the black line, the temperature fluctuates around the setpoint. There is a clear temperature violation in the afternoon when the space changes from unoccupied mode to occupied mode. It takes 4 hours and 45 minutes to cool back down to the setpoint, resulting in a total temperature violation of 14°C-h.

For the LSTM model, shown by the blue line, the controller could not find the optimal solution because the LSTM learned a completely contradictory dynamic response, as mentioned earlier. The controller tends to introduce heating to decrease the space air temperature, which is not feasible during the cooling season, leading to the HVAC being constantly turned off for the LSTM-based dynamic model. Due to the controller failure, the temperature could not be maintained at all, resulting in a total temperature violation of 89°C-h. This highlights the significance of physical consistency in dynamic model development.

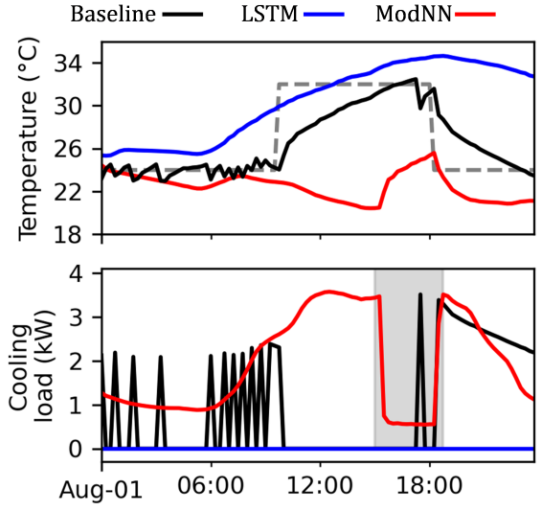

**Figure 11 MPC performance of Baseline, LSTM and ModNN**

The ModNN demonstrates the best control performance, as shown by the red line. The space air temperature remains close to the setpoint to avoid unnecessary energy wastes and a total temperature violation is 0.57°C-h compared to 14°C-h for baseline and 89°C-h for LSTM. Additionally, it achieves up to a 78% peak load reduction due to pre-cooling.

## 4   Discussions

With the development of sensor technology, an increasing amount of data is available in the building sector, providing significant opportunities for data-driven building energy modeling. From a data science perspective, accuracy indicators such as MSE are commonly used to select and compare model performance. However, this is insufficient for control-oriented model, since they should be able to capture the correct response to system inputs. Therefore, we established a new framework to evaluate model performance from both accuracy and consistency perspectives, as shown in Table 3.

**Table 3 Performance indicators for control-oriented building dynamic model**

| Performance indicators | Perspective |
| --- | --- |
| MSE, RMSE, MAE | Accuracy |
| TRV | Consistency (Qualitatively) |
| MMD | Consistency (Quantitively) |

Accuracy is the basic requirement for a data-driven model as it directly affects how closely the prediction results align with the measurement results. In MPC problems, the optimal control decision is calculated based on the predicted system dynamics. Inaccurate prediction results could lead to significant violations of control performance.

TRV can qualitatively describe whether a model responds correctly or not. This indicator represents the underlying physical principles and has to be strictly satisfied. The calculation of TRV can be adjusted based on physical prior knowledge. For example, in this case study, we consider that space air temperature should decrease with more cooling. A tighter statement can be made: space air temperature should not only decrease with more cooling but also remain higher than the minimum value of supply air temperature and outdoor air temperature during the daytime (Law of Thermodynamics). However, stricter constraints require more physical considerations, which increase model complexity and can decrease model accuracy. As shown in Figure 12, physical constraints can shrink the solution space and lead to a drop in accuracy while ensuring consistency. Therefore, the trade-off relationship between accuracy and consistency needs to be explored in future work.

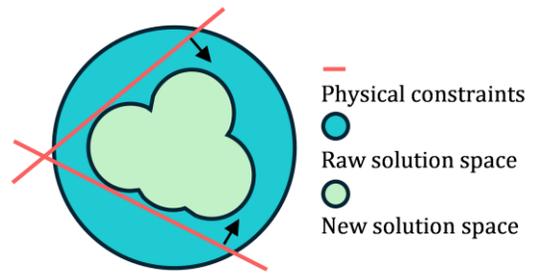

**Figure 12 Physical consistency vs accuracy**

MMD provide a quantitively index for physical consistency evaluation. A value closer to 0 indicates better similarity between the two datasets, meaning the system responses are more closely aligned. Take Figure 13 as example, where the MAE is 0.6°C better than the result from Figure 5 and the model has 0 TRV indicating that the model responds correctly. However, the space air temperature is not sensitive to HVAC power. The temperature barely changes with minimal/maximum cooling supply.

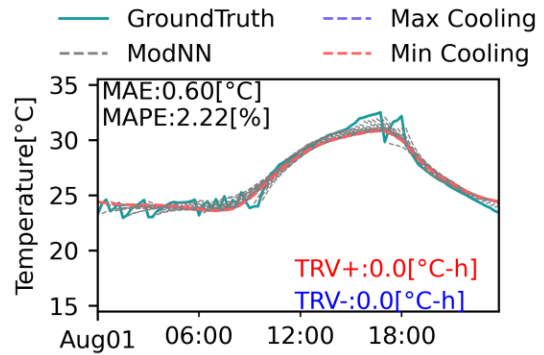

**Figure 13 ModNN with bad MDD performance**

Then we present the MMD result as shown in Figure 14, where the MMD of ModNN is 0.09, higher than the result shown in Figure 10. This case demonstrates that a control-oriented model needs to satisfy accuracy requirements and evaluate physical consistency both qualitatively and quantitatively.

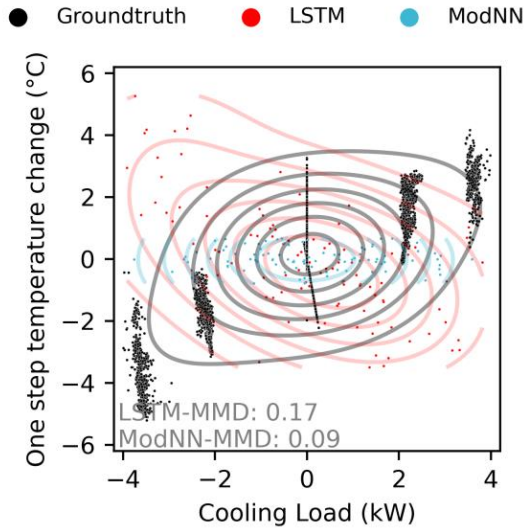

**Figure 14 MDD of ModNN with insensitive model response**

## 5 Limitation and future studies

### 5.1 Customized consistency loss function is needed for model training

We proposed a physical consistency evaluation framework to verify the physical consistency of a given data-driven model. However, the current ModNN can only ensure that the response is qualitatively correct based on model parameter constraints established from Eq. 2. It cannot quantitatively guarantee that the response closely matches real-world system dynamics. For instance, the ModNN might predict a 1°C temperature decrease under a 2-kW cooling load, while the ground truth might only decrease by 0.5°C. In future studies, we want to incorporate MMD loss into our training process to ensure that our model not only responds correctly but also accurately.

### 5.2 Sensitivity analysis of MSE, TRV and MMD

A more comprehensive study should be conducted in future research to evaluate how a model's accuracy and consistency influence control performance and to identify the balance point between model accuracy and consistency.

## 6 Conclusions

In summary, we proposed a modularized encoder–decoder neural network that incorporates heat balance principles for building control optimization. We developed a physical consistency evaluation framework that qualitatively and quantitatively assesses the physical consistency of any given data-driven model. We compared the accuracy, consistency, and control performance of our ModNN and the classical LSTM model using an EnergyPlus-ModNN-MPC co-simulation virtual testbed. The proposed ModNN consistently guarantees physical consistency and achieves significant peak load shifts, whereas the LSTM model, despite its accurate predictions, responds incorrectly and is therefore unsuitable for this energy optimization problem.

We believe that the proposed model and evaluation framework offer a scalable solution for future Building Energy Modeling (BEM) without substantial modeling efforts and customized case-by-case calibrations, paving the way for advanced multi-scale, multi-component, and multi-task building energy modeling and control optimization.

## ACKNOWLEDGMENTS

This work was supported by the U.S. National Science Foundation (Award No. 1949372).